\documentstyle[preprint,aps]{revtex}

\begin{document}
\title{Local polariton states in impure ionic crystals}
\author{V.S. Podolsky, Lev I. Deych and A.A. Lisyansky}
\address{Department of Physics, Queens College, CUNY, Flushing, NY 11367}
\date{\today}
\maketitle

\begin{abstract}
We consider the dynamics of an ionic crystal with a single impurity in the
vicinity of the polariton resonance. We show that if the polariton spectrum
of the host crystal allows for a gap between polariton branches, the defect
gives rise to a novel kind of local states with frequencies within the gap.
Despite the atomic size of the impurity we find that new local states are
predominated by long-wavelength polaritons. The properties of these states are
shown to be different from the properties of the well-known vibrational
local states. The difference is due to the singular behavior of the density
of states of polaritons near the low-frequency boundary of the polariton
gap. Assuming cubic simmetry of the defect site we consider a complete set
of the local states arising near the bottom of the polariton gap.
\end{abstract}

\pacs{71.36.+c,63.20.Pw,63.20.Dj}

\section{Introduction}

An existence of local excitations caused by crystal defects is a well-known
phenomenon in solid state physics \cite{Lifshitz4,Maradudin}. A single
point-like defect can give rise to new states localized in the vicinity of a
defect with frequencies outside the bands of extended states of a host
crystal. These states and their interaction with electromagnetic waves (IR
absorption, Raman scattering) were extensively studied since the early
forties following pioneering works by Lifshitz \cite
{Lifshitz1,Lifshitz2,Lifshitz3}. In all these studies electromagnetic field
was considered as an external field, which excites vibrational states.
The feedback effect of the vibrations upon electromagnetic field was
neglected. However, in ionic crystals in the region of frequencies closed to
the crossing point of phonon and photon dispersion curves one has to take
this effect into account because the coupling between electromagnetic waves
and vibrations becomes so strong that the new kind of excitations --
polaritons -- emerges. In this situation the analyses of local vibrations
and their interaction with electromagnetic field has to be done more
carefully with taking the polariton effects into account. For the first time
such a consideration was carried out in Ref.\cite{Leva} where a novel kind
of local excitations was discovered. It was shown in Ref.\cite{Leva} that if
the polariton spectrum of a crystal exhibits a gap between polariton
branches, where the density of states of a host material is equal to zero, a
defect embedded in the crystal gives rise to local states with frequencies
within the gap. These states are a mix of electromagnetic field and
excitations of a host material. Their properties appeared to be quite
different from those of known local phonon states. Unlike pure phonon
systems, in the case of isotropic materials there is no minimum critical
value of the ``strength'' of the impurity for local polariton states to
appear. This feature was shown to be caused by a negative dispersion of
optical phonons resulting in a non-monotonic dispersion of polaritons.

A general analysis made in Ref.\cite{Leva} refers to arbitrary
``polarization waves.'' It is applicable to a variety of excitations, such
as phonon-polaritons, exciton-polaritons, etc. Concerning a particular case
of phonon-polaritons the model of a dipole interaction between scalar
electromagnetic and polarization waves, used in Ref.\cite{Leva}, needs to be
extended to account for a vector nature of the excitations. Moreover, the
analysis of non-isotope impurities, affecting the elastic bonds around a
defect, is not consistent within the scalar model.

In the present paper we take into account the vector nature of
electromagnetic waves interacting with vibrations in crystals with cubic
symmetry. We assumed that an impurity atom in addition to having a different
mass can locally change elastic constants. The anisotropy of the crystal is
assumed to be weak and is neglected in the long wavelength limit. We
obtained two series of local states which differ in parity. In agreement
with results of Ref.\cite{Leva}, all these states appear first at the bottom
of the polariton gap for infinitesimally small variations of impurity
parameters. This is shown to be caused by a singularity of the density of
states in the lower polariton band. This singularity also provides that the
localization of transverse polaritons is most effective near the lower
boundary of the gap. We show that the local polariton states, unlike usual
transverse extended states are affected by the interaction with the
longitudinal phonon modes. This interaction narrows the frequency range
available for the local polariton states.

\section{Polaritons in a pure crystal: isotropic approximation}

We consider below a body centered cubic (BCC) dielectric crystal with two
oppositely charged ions per each elementary cell. The interaction between
ions is assumed to be a central one. Denoting masses of the positive and
negative ions as $m_{+}$ and $m_{-}$, respectively, and their displacements
as ${\bf U_{\pm}(r)}$, where ${\bf r}$ belongs to the positive or negative
sublattice, we can write down the equations of stationary lattice vibrations
coupled to a coherent electric field, ${\bf E(r)}$: 
\begin{equation}
m_{\pm}\omega^{2}{\bf U^{}_{\pm}(r)}\pm q {\bf E(r)}= -\sum_{s} \left[ \sum_{%
{\bf R}_s} {\bf f} \left({\bf r}\,{\bf R}_s \right)+ \sum_{{\bf R^{\prime}}%
_s} {\bf f^{\prime}}\left({\bf r}\,{\bf R^{\prime}}_s\right) \right],
\end{equation}
where $q$ denotes an ion charge, $\omega$ is the frequency.

The right-hand side of Eq.~(1) presents all elastic forces acting on the ion
hosted at the site ${\bf r.}$ Vectors ${\bf R}_s$ and ${\bf R}^{\prime}_s$
denote radius-vectors of neighboring ions in the $s$-shell spheres of the
original and alternative sublattice, respectively. In the case of a central
interaction the elastic forces within one sublattice have a form: 
\begin{equation}
{\bf f} \left({\bf r}\,{\bf R}_s \right)=-\frac{\beta(R_s)}{R_{s}^{2}} {\bf R%
}_{s}\left\{{\bf R}_{s}\cdot\left[ {\bf U}^{}_{\pm}( {\bf r})-{\bf U}%
^{}_{\pm}({\bf r}\!+\!{\bf R}_s) \right]\right\}
\end{equation}
and forces between ions from different sublattices are: 
$$
{\bf f^{\prime}}\left({\bf r}\,{\bf R^{\prime}}_s\right)=-\frac{%
\beta^{\prime}(R^{\prime}_s)}{R_{s}^{^{\prime}2}} {\bf R^{\prime}}_{s}\left\{%
{\bf R^{\prime}}_{s}\cdot\left[ {\bf U}^{}_{\pm}( {\bf r})-{\bf U}^{}_{\mp}(%
{\bf r}\!+\!{\bf R^{\prime}}_s) \right]\right\},\eqno(2') 
$$
where elastic constants of intra- and inter-sublattice interaction, $%
\beta(R) $ \ and \ $\beta^{\prime}(R^{\prime}),$ depend on a distance
between ions only.

A coherent electric field induced by the ionic vibrations, ${\bf E(r)},$
invokes an additional pair of equations: 
\begin{equation}
\nabla\cdot{\bf E(r)}\!=-4\pi q\sum_{{\bf l}} \left[{\bf U_{{}_{+}}(l)}\cdot
\nabla\delta\left({\bf r\!-\!l}\right)\!-\! {\bf U_{{}_{-}}(l\!+\!b)}\cdot
\nabla\delta\left({\bf r\!-\!l\!-\!b}\right)\right],
\end{equation}
\begin{equation}
\nabla\times\left[\nabla\times{\bf E(r)}\right]\!+\!\frac{\omega^{2}}{c^2}%
{\bf E(r)}\!= -\frac{4\pi q\omega^{2}}{c^2}\sum_{{\bf l}} \left[{\bf %
U_{{}_{+}}(l)} \delta\left({\bf r\!-\!l}\right)\!-\! {\bf U_{{}_{-}}(l\!+\!b)%
} \delta\left({\bf r\!-\!l\!-\!b}\right)\right],
\end{equation}
where vectors ${\bf l}$ denote lattice vectors, ${\bf b}$ is the basis
vector and $c$ is the speed of light.

The lattice normal modes arise as simultaneous solutions of Eqs.~(1,3,4).
Fourier transformation of this system gives the dynamics equations in the
momentum representation: 
\begin{equation}
\left( 
\begin{array}{cc}
{{\bf \hat{D}(k)+\hat{F}(k)}-\omega^{2}m_{+}} & {{\bf \hat{D^{\prime}}(k)-%
\hat{F}(k)}} \\ 
{{\bf \hat{D^{\prime}}(k)-\hat{F}(k)}} & {{\bf \hat{D}(k)+\hat{F}(k)}%
-\omega^{2}m_{-}}
\end{array}
\right) \left( 
\begin{array}{c}
{\bf A_{+}(k)} \\ 
{\bf A_{-}(k)}
\end{array}
\right)=0.
\end{equation}
Here ${\bf A_{\pm}(k)}$ denote the Fourier amplitudes of the displacements, $%
{\bf \hat{D}(k)}$ is the dynamical matrix of the intra-sublattice
interaction: 
\begin{equation}
{\bf \hat{D}(k)}\!=\!\sum_{s}\frac{\beta(R_s)}{R_{s}^{2}} \sum_{{\bf R}_s} 
{\bf R}_{s}\!\otimes\!{\bf R}_{s} \left(1\!-\!%
\mbox{\large$e^{i{\bf k\cdot
R}_s}$}\right)\!+\! \frac{\beta^{\prime}(R^{\prime}_s)}{R_{s}^{^{\prime}2}}
\sum_{{\bf R^{\prime}_s}} {\bf R^{\prime}}_{s}\!\otimes\!{\bf R^{\prime}}_{s}
\end{equation}
and ${\bf \hat{D^{\prime}}(k)}$ is the dynamical matrix of the
inter-sublattice interaction: 
$$
{\bf \hat{D^{\prime}}(k)}=-\sum_{s}\frac{\beta^{\prime}(R^{\prime}_s)}{%
R_{s}^{^{\prime}2}} \sum_{{\bf R^{\prime}}_s}{\bf R^{\prime}}_{s}\!\otimes\! 
{\bf R^{\prime}}_{s} \mbox{\large$e^{i{\bf k\cdot R'}_s}$},\eqno(6') 
$$
where ${\bf R\!\otimes\!R}$ denote the direct products of two vectors.

The Fourier amplitudes of the field induced by the lattice vibrations can be
expressed as 
\mbox{${\bf
E}\!=\!-{\bf\hat{F}\left(\bf A_{{}_{+}}\!-\!\bf A_{{}_{-}}\right)}/q,$} with
operator ${\bf \hat{F}}$ defined as follows: 
\begin{equation}
{\bf \hat{F}(k},\omega)\!= \frac{4\pi q^{2}}{a^3}\cdot\frac{\omega^2{\bf 
\hat{I}\!-\!c^2k\!\otimes\!k}}{\omega^2\!-\!c^2k^2} \!=\!\frac{4\pi q^{2}}{%
a^3}\left({\bf \hat{P}}^{}_{\Vert}\! +\!\frac{\omega^2}{\omega^2\!-\!c^2 k^2}%
{\bf \hat{P}}^{}_{\bot}\right),
\end{equation}
where ${\bf \hat{I}}$ is the unit tensor, ${\bf P}^{\alpha\beta}_{\Vert}=k_{%
\alpha}k_{\beta}/k^2$ and ${\bf {P}}^{\alpha\beta}_{\bot}=\delta_{\alpha%
\beta}-k_{\alpha}k_{\beta}/k^2$ are the longitudinal and transverse
projectors, respectively.

The dynamical matrices given by Eqs.~(6,6') are of a general type and
solutions of Eq.~(5), in general, do not split into longitudinal and
transverse modes. However, considering long-wave excitations in weakly
anisotropic crystals we can make use of the isotropic approximation for the
dynamical matrices: 
$$
{\bf \hat{D}(k)}=\gamma _{\Vert }^{{}}(k){\bf \hat{P}}_{\Vert }^{{}}+\gamma
_{\bot }^{{}}(k){\bf \hat{P}}_{\bot }^{{}},\eqno(8) 
$$
$$
{\bf \hat{D^{\prime }}(k)}=\gamma _{\Vert }^{\prime }(k){\bf \hat{P}}_{\Vert
}^{{}}+\gamma _{\bot }^{\prime }(k){\bf \hat{P}}_{\bot }^{{}},\eqno(8') 
$$
where scalar functions, $\gamma _{\sigma }^{{}}(k)$ and $\gamma _{\sigma
}^{\prime }(k),$ can be expressed in terms of frequencies of longitudinal
and transverse phonons. In a crystal of cubic symmetry the dynamical
matrices become trivial at the center of the Brillouin zone: %
\setcounter{equation}{8} 
\begin{equation}
{\bf \hat{D}(0)}\!=-{\bf \hat{D^{\prime }}(0)}\!=\!\sum_{s}\frac{\beta
^{\prime }(R_{s}^{\prime })}{R_{s}^{^{\prime }2}}\sum_{{\bf R_{s}^{\prime }}}%
{\bf R_{s}^{\prime }\!\otimes \!R_{s}^{\prime }}\!=\!\frac{1}{3}{\bf \hat{I}}%
\sum_{s}\beta ^{\prime }(R_{s}^{\prime })Z_{s}^{\prime },
\end{equation}
where $Z_{s}^{\prime }$ denotes a total number of ions in $s$-shell and we
make use of the identity, 
\mbox{$\sum_{\bf R'_s}{\bf  R'_s \otimes
R'_s}\!=\!{\bf\hat{I}}Z'_{s}R^{'2}_s/3$,} valid in crystals with cubic
symmetry. Eq.~(9) sets a condition for parameters in Eqs.~(8), so that all $%
\gamma _{\sigma }^{{}}(0)\!$ and $-\gamma _{\sigma }^{\prime }(0)$ are equal
to a positive constant $\gamma \!=\!1/3\sum_{s}\beta ^{\prime
}(R_{s}^{\prime })Z_{s}^{\prime }.$

In the isotopic approximation all normal modes of a crystal become either
longitudinal or transverse, ${\bf A^{(\sigma)}_{\pm}(k)\!=\!e_{\sigma}(k)}%
A^{(\sigma)}_{\pm}({\bf k}),$ where ${\bf e_{\sigma}(k)}$ are longitudinal
or transverse unit polarization vectors, ${\sigma}$ is a polarization index.
Eqs.~(5,8) give the following relation between the Fourier amplitudes of
displacements in different sublattices: 
\begin{equation}
A^{(\sigma)}_{+}=\frac{f^{}_{\sigma}\!-\!\gamma^{\prime}_{\sigma}} {%
\gamma_{\sigma}+f^{}_{\sigma}\!-\!\omega^{2}m_{+}} A^{(\sigma)}_{-}.
\end{equation}
For longitudinal modes $f^{}_{\Vert}\!=\!4\pi q^{2}/{a^3}\!=\!f$. The
corresponding dispersion equation, 
\begin{equation}
\left(\frac{\gamma_{\Vert} \!+\!f_{\Vert}}{m_{+}}\!-\omega^{2}\right) \left(%
\frac{\gamma_{\Vert}\!+\!f_{\Vert}}{m_{-}} \!-\omega^{2}\right)\!- \frac{%
\left(\gamma^{\prime}_{\Vert}\!-\!f_{\Vert} \right)^2}{m_{+}m_{-}}=
\left(\omega^{2}\!-\!\omega^{2}_{+\Vert}\right)\!
\left(\omega^{2}\!-\!\omega^{2}_{-\Vert}\right)=0,
\end{equation}
defines acoustic, $\omega^{2}_{\Vert}(k),$ and optical, $\Omega^{2}_{%
\Vert}(k),$ branches of longitudinal phonons. For transverse modes $%
f^{}_{\bot}\!=\!f\omega^2/\left(\omega^2\!-\!c^2 k^2\right),$ therefore, the
dispersion equation, 
\begin{equation}
\left(\frac{\gamma_{\bot}\!+\!f_{\bot}}{m_{+}}\!-\omega^{2}\!\right)\! \left(%
\frac{\gamma_{\bot}\!+\!f_{\bot}}{m_{-}}\!-\omega^{2}\!\right)\!- \frac{%
\left(\gamma^{\prime}_{\bot}\!-\!f_{\bot}\right)^2}{m_{+}m_{-}}\!= \frac{%
\left(\omega^{2}\!-\!\omega^{2}_{\bot}\right)\!
\left(\omega^{2}\!-\!\Omega^{2}_{+}\right)\!
\left(\omega^{2}\!-\!\Omega^{2}_{-}\right)}{\omega^2\!-\!c^2 k^2}=\!0,
\end{equation}
gives one acoustic, $\omega^{2}_{\bot}(k),$ and two polariton, $%
\Omega^{2}_{\pm}(k),$ branches.

Neglecting the field effects and solving Eqs.~(11,12) in the long wavelength
limit one can obtain expressions for $\gamma _{\sigma }(k)$ and $\gamma
_{\sigma }^{\prime }(k)$ in terms of conventional parameters: 
\begin{equation}
\gamma _{\sigma }(k)\approx \gamma \!+\!\mu k^{2}\left( v_{\sigma
}^{2}\!-\!v_{\sigma }^{^{\prime }2}\right) ,
\end{equation}
$$
\gamma _{\sigma }^{\prime }(k)\approx -\gamma \!+\!\mu k^{2}\left[ v_{\sigma
}^{^{\prime }2}\!+\!v_{\sigma }^{2}\left( \frac{M}{2\mu }\!-\!1\right)
\right] ,\eqno(13') 
$$
where $\mu $ and $M$ are reduced and total masses of ions within the
elementary cell, respectively, $v_{\sigma }$ and $v_{\sigma }^{\prime }$ are
the velocities of acoustic and optical phonons with a given polarization.

Eqs.~(11,12) show that the internal field affects dispersion relations of
all lattice excitations. However, the physical effects of the photon-phonon
interaction are substantial in a vicinity of the polariton resonance which
takes place in the long wavelength region. It is straightforward to show
that the acoustic branches, 
\begin{equation}
\omega _{\sigma }^{2}(k)\approx 2\frac{\gamma _{\sigma }+\gamma _{\sigma
}^{\prime }}{M}\approx v_{\sigma }^{2}k^{2},
\end{equation}
are unaffected by the field, whereas, the interaction with the field results
in the uniform up-shift of the longitudinal optical branch, 
\begin{equation}
\omega ^{2}\!=\!\Omega _{\Vert }^{2}(k)\!+\!d^{2},
\end{equation}
and leads to the well-known polariton dispersion relation for optical
transverse excitations \cite{Born,Kittel}, 
\begin{equation}
\left[ \omega ^{2}\!-\!\Omega _{\bot }^{2}(k)\right] \left[ \omega
^{2}\!-\!c^{2}k^{2}\right] \!=\!d^{2}\omega ^{2}.
\end{equation}
Eq.~(16) describes two polariton branches, $\Omega _{\pm }(k)$, with
corresponding dispersion laws: 
\begin{equation}
\Omega _{\pm }^{2}(k)\!=1/4\left( \sqrt{\left[ \Omega _{\bot
}(k)\!+\!ck\right] ^{2}\!+\!d^{2}}\!\pm \!\sqrt{\left[ \Omega _{\bot
}(k)\!-\!ck\right] ^{2}\!+\!d^{2}}\right) ^{2},
\end{equation}
where $\Omega _{\bot }$ and $\Omega _{\Vert }$ are frequencies of the
transverse and longitudinal optical phonons, respectively, 
\begin{equation}
\Omega _{\sigma }^{2}(k)\approx \frac{\gamma _{\sigma }}{\mu }-2\frac{\gamma
_{\sigma }+\gamma _{\sigma }^{\prime }}{M}\,\approx \omega
_{0}^{2}-v_{\sigma }^{^{\prime }2}k^{2},
\end{equation}
a phonon-photon coupling parameter, $d^{2}\!=\!f/\mu \!=\!4\pi q^{2}/\mu
a^{3},$ is the ionic ``plasma frequency'' and $\omega _{0}^{2}\!=\!\gamma
/\mu $ is the optical activation frequency.

Analysis of Eqs.~(14,15,17) shows that the lower polariton branch, $%
\Omega^{2}_{-}(k),$ has a zero activation frequency and extends over both
acoustic bands, whereas, the longitudinal optical branch, $%
\Omega^{2}_{\Vert}(k)\!+\!d^2,$ overlaps with the top part of the polariton
gap. Therefore, a truly forbidden gap, with no modes of any kind inside, may
exist only between the lower boundary of the polariton gap and the bottom of
the longitudinal band [Fig.~1]. When the dispersion of optical phonons is
neglected, both polariton branches are monotonic and the spectral gap is
between the transverse and longitudinal frequencies, $\omega_0^2$ \ and \ $%
\omega_0^2+d^2.$

It turns out that accounting for the phonon dispersion causes a {\it %
qualitative} change of this picture. First, the upper boundary of the
frequency gap, $\omega_{2}^{2},$ is now set by the bottom of the optical
longitudinal band. Second, the lower polariton branch in the case of a
negative dispersion becomes non-monotonic and gains a maximum at some $%
k\!=\!k_0,$ close to the center of the Brillouin zone [Fig 1.]. Calculations
in the long wavelength approximation give $k_0^2\approx
2\omega_0d/v^{\prime}_{\bot}c$ and a new lower boundary of the gap, $%
\omega^{2}_{1}\!=\!\Omega^2_{-}(k_0)\approx\omega_0^2\!-2v^{\prime}_{\bot}%
\omega_0d/c.$ Since $2v^{\prime}_{\bot}\omega_0d/c\approx
v_{\bot}^{^{\prime}2}k_0^2\ll\omega_0^2,$ the result is consistent with the
approximation. However, because $c^2k_0^2\approx
2c\omega_0d/v^{\prime}_{\bot}\gg\omega_0^2,$ \ the maximum of $%
\Omega^2_{-}(k)$ is far away from the very narrow (due to $c\gg
v^{\prime}_{\bot}$) polariton resonance region. Since $\Omega^{2}_{-}(k)$
reaches its maximum at the surface of a finite area inside the Brillouin
zone, the density of polariton states diverges at the gap's lower boundary.
As we show latter, it causes an absence of a lower threshold for local
polariton states.

\section{polariton local states}

When a host ion at the site ${\bf r\!=\!0}$ of the positive sublattice is
replaced by an impurity ion with the same charge, it causes a local
deviation of the crystal density and a local change of elastic constants. To
account for these facts we need to add extra forces to the right hand side
of Eq.~(1). For positive and negative sublattices, respectively, these
forces are: 
\begin{equation}
\delta {\bf f^{+}(r)}\!=\!\left\{ \omega ^{2}\delta m{\bf U_{+}^{{}}(0)}-%
\frac{\delta \beta }{n^{2}}\sum_{{\bf n}}{\bf n}\left[ {\bf n\cdot
U_{+}^{{}}(0)\!-\!n\cdot U_{-}^{{}}(n)}\right] \right\} \delta _{{\bf r\,0}}
\end{equation}
and 
$$
\delta {\bf f^{-}(r)}\!=\!-\frac{\delta \beta }{n^{2}}\sum_{{\bf n}}{\bf n}%
\left[ {\bf n\cdot U_{+}^{{}}(0)\!-\!n\cdot U_{-}^{{}}(n)}\right] \delta _{%
{\bf r\,n}},\eqno(19') 
$$
where $\delta m$ is the difference between the masses of an impurity and a
host ion, $\delta \beta $ is a shift in the elastic constant in the
impurity's near-neighbor shell and vectors ${\bf n}$ denote radius-vectors
of the impurity's nearest neighbors.

The dynamic equation (5) is now modified: 
\begin{equation}
\left( 
\begin{array}{cc}
{{\bf \hat{D}(k)+\hat{F}(k)}-\omega^{2}m_{+}} & {{\bf \hat{D^{\prime}}(k)-%
\hat{F}(k)}} \\ 
{{\bf \hat{D^{\prime}}(k)-\hat{F}(k)}} & {{\bf \hat{D}(k)+\hat{F}(k)}%
-\omega^{2}m_{-}}
\end{array}
\right) \left( 
\begin{array}{c}
{\bf A_{+}(k)} \\ 
{\bf A_{-}(k)}
\end{array}
\right)= \frac{1}{N} \left( 
\begin{array}{c}
{\bf B_{+}} \\ 
{\bf B_{-}}
\end{array}
\right),
\end{equation}
where $N$ is the number of ions in one sublattice and 
\begin{equation}
{\bf B_{+}}\!=\! \omega^2 \delta m{\bf U^{}_{+}} \!-\frac{\delta\beta}{n^2}%
\sum_{{\bf n}}{\bf n}\! \left[{\bf n\cdot U^{}_{+}}\!-\!U^{}_{{\bf n}%
}\right],
\end{equation}
$$
{\bf B_{-}}\!=\!\frac{\delta\beta}{n^2}\sum_{{\bf n}}{\bf n} \left[{\bf %
n\cdot U^{}_{+}}-U^{}_{{\bf n}}\right]\mbox{\large$e^{-i\bf k\cdot n}$}.%
\eqno(21') 
$$
Here ${\bf U^{}_{+}\!=\!U^{}_{+}(0)}$ and all ${\bf U^{}_{-}(n)}$ appear
only in combinations $U^{}_{{\bf n}}\!=\!{\bf n \cdot U^{}_{-}(n)}$, because
ion-ion forces are central.

Let us denote the matrix in the left-hand side of Eq.~(20) as 
Cartesian ($\alpha $) and sublattice ($\varepsilon $) indices of the
amplitudes $A_{\varepsilon }^{\alpha }$. In the isotropic approximation this
matrix can be decomposed, 
\mbox{${\mbox{\large$\bf L$}}=
{\bf L}^{}_{\Vert}\otimes{\bf\hat{P}}^{}_{\Vert}+
{\bf L}^{}_{\bot}\otimes{\bf\hat{P}}^{}_{\bot},$}\ where $2\times 2$%
-matrices ${\bf L}_{\sigma }^{{}}$, operating on sublattice indexes only,
are defined as the follows: 
\begin{equation}
{\bf L}_{\sigma }^{{}}=\left( 
\begin{array}{cc}
{l_{\sigma }^{+}} & {l_{\sigma }^{\prime }} \\ 
{l_{\sigma }^{\prime }} & {l_{\sigma }^{-}}
\end{array}
\right) =\left( 
\begin{array}{cc}
{\gamma _{\sigma }+f_{\sigma }-\omega ^{2}m_{+}^{{}}} & {\gamma _{\sigma
}^{\prime }-f_{\sigma }} \\ 
{\gamma _{\sigma }^{\prime }-f_{\sigma }} & {\gamma _{\sigma }+f_{\sigma
}-\omega ^{2}m_{-}^{{}}}
\end{array}
\right) .
\end{equation}
The inverse matrix, ${\mbox{\large$\bf L$}}^{-1}\!={\bf L}_{\Vert
}^{-1}\otimes {\bf \hat{P}}_{\Vert }^{{}}+{\bf L}_{\bot }^{-1}\otimes {\bf 
\hat{P}}_{\bot }^{{}}={\mbox{\large$\bf G$}}={\bf G}_{\Vert }^{{}}\otimes 
{\bf \hat{P}}_{\Vert }^{{}}+{\bf G}_{\bot }^{{}}\otimes {\bf \hat{P}}_{\bot
}^{{}},$ where 
\begin{equation}
{\bf G}_{\sigma }\!=\!\left( 
\begin{array}{cc}
{g_{\sigma }^{+}} & {g_{\sigma }^{\prime }} \\ 
{g_{\sigma }^{\prime }} & {g_{\sigma }^{-}}
\end{array}
\right) \!\!=\!\frac{1}{m_{+}m_{-}}\!\!\left[ \!\left( \frac{\gamma _{\sigma
}\!+\!f_{\sigma }}{m_{+}}\!-\!\omega ^{2}\right) \!\!\left( \frac{\gamma
_{\sigma }\!+\!f_{\sigma }}{m_{-}}\!-\!\omega ^{2}\right) \!\!-\!\frac{%
\left( \gamma _{\sigma }^{\prime }\!-\!f_{\sigma }\right) ^{2}}{m_{+}m_{-}}%
\right] ^{-1}\!\!\left( 
\begin{array}{cc}
{l_{\sigma }^{-}} & {-l_{\sigma }^{\prime }} \\ 
{-l_{\sigma }^{\prime }} & {l_{\sigma }^{+}}
\end{array}
\right) ,
\end{equation}
is the Green's function of the system in the momentum representation. In
accordance with a general property of Green's functions, poles of its matrix
elements, $g_{\sigma }(\omega ,{\bf k})$, coincide with eigenfrequencies of
a pure crystal, 
\[
g_{\Vert }(\omega )\propto \left( \omega ^{2}\!-\!\omega _{\Vert
}^{2}\right) ^{-1}\left( \omega ^{2}\!-\!\Omega _{\Vert
}^{2}\!-\!d^{2}\right) ^{-1}, 
\]
\[
g_{\bot }(\omega )\propto \left( \omega ^{2}\!-\!\omega _{\bot }^{2}\right)
^{-1}\!\left( \omega ^{2}\!-\!\Omega _{+}^{2}\right) ^{-1}\!\left( \omega
^{2}\!-\!\Omega _{-}^{2}\right) ^{-1}, 
\]
as it follows from Eqs.~(11,12).

Solving Eq.~(20) with the help of Eqs.~(22,23), one can obtain Fourier
amplitudes of the ion displacements, ${\bf A_{+}(k)}$ and ${\bf A_{-}(k)}$,
and the displacements themselves: 
$$
{\bf U_{+}(r)}=\sum_{\sigma }^{{}}\int_{({\bf k)}}%
\mbox{\large$e^{i\bf
k\cdot r}$}\,{\bf e_{\sigma }(k)\!\otimes \!e_{\sigma }(k)}\left( g_{\sigma
}^{+}{\bf B_{+}^{{}}}+g_{\sigma }^{\prime }{\bf B_{-}^{{}}},\right) \eqno(24)
$$
$$
{\bf U_{-}(r)}=\sum_{\sigma }^{{}}\int_{({\bf k)}}%
\mbox{\large$e^{i\bf
k\cdot r}$}\,{\bf e_{\sigma }(k)\!\otimes \!e_{\sigma }(k)}\left( g_{\sigma
}^{\prime }{\bf B_{+}^{{}}}+g_{\sigma }^{-}{\bf B_{-}^{{}}},\right) %
\eqno(24') 
$$
where a symbol $\int_{({\bf k)}}$ denotes $\left( 1/N\right) \sum_{{\bf k}%
}\approx \left( a/2\pi \right) ^{3}\int d{\bf k},$ with wave vectors taken
from the first Brillouin zone only.

Substitution of ${\bf B_{+}}$ and ${\bf B_{-}}$, given by Eqs.~(21,21'),
into Eqs.~(24) allows one to express all displacements in terms of ${\bf %
U^{}_{+}} $ and $U^{}_{{\bf n}}$ only, and then obtain a closed system of
equations for these variables. In the case of an {\it isotope impurity},
considered earlier in the scalar model of Ref.~\cite{Leva}, we set $%
\delta\beta\!=\!0$ and this system of {\it spectral equations} reads: %
\setcounter{equation}{24} 
\begin{equation}
{\bf U_{+}}=\omega^2\delta m \sum^{}_{\sigma}\int_{({\bf k)}}g^{+}_{\sigma}\,%
{\bf e_{\sigma}(k)\!\otimes\!e_{\sigma}(k)}{\bf U^{}_{+}}.
\end{equation}
In a cubic crystal, vector ${\bf U_{+}}$ can be arbitrary [Appendix A],
whereas frequencies of excitations are defined by the equation: 
\begin{equation}
1=\frac{\omega^2\delta m}{3} \int_{({\bf k)}}\left(g^{+}_{\Vert}\!+%
\!2g^{+}_{\bot}\right)=\delta m \omega^2 I(\omega^2),
\end{equation}
which generalizes Eq.~(10) obtained in Ref.\cite{Leva}.

All solutions of these equations can be divided into two classes: extended
and local states. Since an impurity destroys the translational symmetry of
the crystal, any state is now a superposition of all normal modes available
in the first Brillouin zone. For extended states, corresponding to
scattering states with well-defined wave vectors, frequencies fall into the
bands of a pure crystal. Local states, dependent upon the value of the
parameter $\delta m$ in Eq.~(26), may arise outside of the bands. From the
structure of $g_{\sigma }(\omega ,{\bf k})$ it might seem that when a
frequency is close to a particular band, the modes from it dominate in the
corresponding state. However, because contributions of the near and distant
bands could be weakened or strengthened by the low or high density of states
in them, direct calculations are required here.

As we already mentioned, the function $\Omega^{2}_{-}(k)$ reaches its
maximum value, $\omega^{2}_{1}$,\ at \ $k\!=\!k_{0},$ close to the center of
the Brillouin zone. Its expansion around $k_{0}$ does not contain a linear
term: 
\begin{equation}
\Omega^{2}_{-}(k)\!=\!\omega^{2}_{1}\!-\!\nu^2 a^2\left(k\!-\!k_0\right)^2
\!+\! \mbox{$\cal O$}\left[a^3\left(k\!-\!k_0\right)^3\right].
\end{equation}
It immediately shows that the integral in the left-hand side of Eq.~(26)
diverges at $\omega\!=\!\omega_1$ due to a contribution from the lower
polaritons branch. Therefore, considering $\omega$ in the frequency gap
close to its lower boundary we can omit $g^{+}_{\Vert}$ from Eq.~(26) and
rewrite $\int_{({\bf k)}}g^{+}_{\bot}$ using the density of states, $%
\rho_{-}(\varepsilon)$, in the lower polaritons band: 
\begin{equation}
I^{+}(\omega^2)\!=\!\int_{({\bf k)}}\frac{2g^{+}_{\bot}}{3}\!=\! \left(\frac{%
a }{2\pi}\right)^3\int \frac{2\left(\omega^2\!-\!c^2 k^2\right)
\left(\gamma_{\bot}+f_{\bot}-\omega^{2}m^{}_{-}\right)d{\bf k}}{3
m_{+}m_{-}\left[\omega^{2}\!-\!\omega^{2}_{\bot}\right]\!
\left[\omega^{2}\!-\!\Omega^{2}_{+}\right]\!
\left[\omega^{2}\!-\!\Omega^{2}_{-}\right]}\!=\! \frac{2}{3}\int_{C}^{}\frac{%
F(k)\rho_{-}(\varepsilon)d\varepsilon} {\omega^2\!-\!\varepsilon},
\end{equation}
where $F(k)$ denotes all factors of $g^{+}_{\bot},$ other than $%
\left(\omega^{2}\!-\!\Omega^{2}_{-}\right)^{-1}$ and the last integration is
performed on a complex $\varepsilon$-plane [Appendix B].

Near the bottom of the polariton band, for small $\varepsilon,$ the density
of states has the usual Kohn shape, $\rho_{-}(\varepsilon)\propto\sqrt{%
\varepsilon}.$ \ The asymptote of $\rho_{-}(\varepsilon)$ near the top of
the band, for small $z\!=\!\omega^{2}_{1}\!-\!\varepsilon,$ can be found
with the help of Eq.~(27): 
\begin{equation}
\rho_{-}(\varepsilon)\!=\!\left(\frac{a}{2\pi}\right)^3\int\frac{ds}{%
|\nabla\Omega^{2}_{-}(k)|} \approx \left(\frac{a}{2\pi}\right)^2 \frac{%
\left(k_{0}\!+\!z^{1/2}/a\nu\right)^2\!+
\!\left(k_{0}\!-\!z^{1/2}/a\nu\right)^2}{\nu z^{1/2}} =\frac{(ak_0)^2}{%
2\pi^{2}\nu z^{1/2}},
\end{equation}
where all omitted terms are regular at $z\!=\!0.$

Since for $\omega \!\rightarrow \!\omega _{1}\!+\!0$ the divergent part of $%
I^{+}(\omega ^{2})$ is provided by a region of small $z,$ in Eq.~(28) we can
replace the exact density of states with the found asymptote. Considering $%
\omega ^{2}\!-\!\omega _{1}^{2}\ll \omega _{1}^{2}$ it allows us to
calculate the leading part of $I^{+}(\omega ^{2})$ and transform the
spectral equation (26) into the form [Appendix B]: 
\begin{equation}
\frac{m_{+}}{\delta m}\approx -\frac{2\left( ak_{0}\right) ^{2}\omega
_{0}^{2}}{3\pi \nu \sqrt{\omega ^{2}\!-\!\omega _{1}^{2}}}\frac{m_{-}}{M}.
\end{equation}
Using estimates made in section.~2 and recalling that \ $a\nu \!=\!v_{\bot
}^{\prime },$ we obtain: 
\begin{equation}
\frac{\delta m}{m}\!=-\frac{3\pi cv_{\bot }^{^{\prime }2}\sqrt{\omega
^{2}\!-\!\omega _{1}^{2}}}{a^{3}\omega _{0}^{3}d}.
\end{equation}
This result recovers Eq.~(14) of Ref.\cite{Leva} obtained in the scalar
model and supports the conclusion of the absence of the lower localization
threshold for light isotope impurities.

In a general case of {\it non-isotope} impurity the spectral system is of
the 11th rank with 11 variables, ${\bf U_{+}\!=\!U_{+}(0)},\,U_{{\bf n}
}^{{}}\!=\!{\bf n\cdot U_{-}^{{}}(n)}$. The system can be further simplified
with the help of the crystal symmetry. The exact point group of a cubic
crystal includes the space inversion. Therefore, all excitations can be
classified by their spatial parity.

For the {\it odd states}, where ${\bf U_{\pm }^{{}}(-r)\!=-U_{\pm }^{{}}(r)}$%
, one can see that both ${\bf U_{+}^{{}}}$ and \ ${\bf B_{+}^{{}}}$ are
equal to zero and the rank of the spectral system reduces from 11 to 4.
Considering displacements of four negative ions at corners of one face of
the BCC-lattice elementary cell (see Fig.~2) as independent variables, from
Eqs.~(24) we obtain the following system: 
\begin{equation}
\sum_{s^{\prime }\!=\!1}^{4}M_{ss^{\prime }}U_{s^{\prime }}\!=-\,\frac{n^{2}%
}{2\delta \beta }\,U_{s},
\end{equation}
where index $s$ numerates the chosen ions, their radius-vectors are denoted
as ${\bf n}_{s}$ and $U_{s}\!=\!U_{{\bf n}_{s}}\!=\!{\bf n}_{s}\cdot {\bf U}%
_{-}^{{}}({\bf n}_{s}).$ The matrix $M_{ss^{\prime }}$ here is given by the
equation: 
\begin{equation}
M_{ss^{\prime }}=M({\bf n}_{s},\,{\bf n}_{s^{\prime }})=\sum_{\sigma
}^{{}}\int_{({\bf k)}}^{{}}g_{\sigma }^{-}({\bf e_{\sigma }\cdot n}_{s})(%
{\bf e_{\sigma }\cdot n}_{s^{\prime }})\sin ({{\bf k\cdot n}_{s}})\sin ({%
{\bf k\cdot n}_{s^{\prime }}}).
\end{equation}
Using transformation properties of the integrand one can show that the
matrix $M_{ss^{\prime }}$ remains invariant, $M({\bf \hat{Q}}{\bf n},\,{\bf 
\hat{Q}}{\bf n^{\prime }})=M({\bf n,\,n^{\prime }}),$ under any point
transformation ${\bf \hat{Q}}$. Therefore, the elements of the symmetric
matrix $M_{ss^{\prime }}$ are equal for equivalent pairs of ions: $%
M_{11}\!=\!M_{22}\!=\!M_{33}\!=\!M_{44},$ \ then \ $M_{12}\!=\!M_{23}\!=%
\!M_{34}$ \ and \ $M_{13}\!=\!M_{24}.$ After finding eigenvalues of this
matrix we obtain three spectral equations: 
\begin{equation}
-\frac{n^{2}}{2\delta \beta }\!=\!\mu _{0}\!=\!M_{11}-M_{13},
\end{equation}
$$
-\frac{n^{2}}{2\delta \beta }\!=\!\mu _{\pm }\!=\!M_{11}+M_{11}\pm 2M_{12}.%
\eqno(34') 
$$
The structure of the eigenvectors shows that the states corresponding to $%
\mu _{+}$ and $\mu _{-}$ represent ``rhombic'' and ``tetragonal''
oscillations localized around the stationary impurity, whereas, the states
corresponding to $\mu _{0}$ involve both types of deformations of the
elementary cell [Appendix C].

Near the lower boundary of the gap the arguments used in evaluation of
Eq.~(26) are also applicable. Retaining the leading terms in Eq.~(33) and
using symmetry properties of the arising expressions [Appendix C], we obtain
the following expressions: 
\begin{equation}
\mu _{0}\!\approx \frac{1}{5}\left[ n^{4}\!-\!({\bf n}_{1}\cdot {\bf n}%
_{3})^{2}\right] \int_{({\bf k)}}^{{}}k^{2}g_{\bot }^{-},
\end{equation}
$$
\mu _{\pm }\!\approx \frac{1}{5}\left\{ \!({\bf n}_{1}\cdot {\bf n}%
_{3})^{2}\!+\!\frac{n^{4}}{3}\pm 2\left[ ({\bf n}_{1}\cdot {\bf n}%
_{2})^{2}\!-\!\frac{n^{4}}{3}\right] \right\} \int_{({\bf k)}%
}^{{}}k^{2}g_{\bot }^{-}.\eqno(35') 
$$
From the geometry of the elementary cell it follows that $({\bf n}_{1}\cdot 
{\bf n}_{2})\!=\!-({\bf n}_{1}\cdot {\bf n}_{3})\!=\!n^{2}/3\!=\!a^{2}/4$
and, therefore, $\mu _{0}=\!\mu _{-}=\!\left( a^{4}/10\right) \int_{({\bf k)}%
}^{{}}k^{2}g_{\bot }^{-}.$ It leads to the unique spectral equation for all
``tetragonal'' modes [Appendix B]: 
\begin{equation}
\frac{\delta \beta }{\gamma }\approx \frac{15\pi \nu \sqrt{\omega
^{2}\!-\!\omega _{1}^{2}}}{4\left( ak_{0}\right) ^{4}\omega _{0}^{2}}\left( 
\frac{M}{m_{+}}\right) ^{2}.
\end{equation}
This result shows that the odd local states arise upon an infinitesimally
small strengthening of local bonds associated with the impurity, $\delta
\beta \ge +0.$ This effect, similar to the isotope impurity case, is due to
a singularity of the density of states in the lower polariton band.
Eq.~(36), compared to Eq.~(30), has an additional small factor, $\left(
ak_{0}\right) ^{2}.$ Therefore, for the same relative deviations, $\delta
\beta /\beta $ and $\delta m/m,$ the odd states lie much closer to the gap's
boundary than the states associated with an isotope impurity. Moreover,
these states are much less sensitive to variations of parameters than the
isotope-induced local states. This fact reveals itself in an extreme form
for ``rhombic'' modes corresponding to $\mu _{+}.$ Eq.~($35^{\prime }$)
gives $\mu _{+}\!=\!0$ which means that within the approximations used these
states appear right at the gap's bottom for any value of $\delta \beta .$
Accounting for the higher-order terms in expansions of the $\sin {}$%
-functions in Eq.~(31) will separate these modes from the gap's boundary but
the separation, $\sqrt{\omega ^{2}\!-\!\omega _{1}^{2}}\propto \delta \beta
\left( ak_{0}\right) ^{6},$ remains the smallest among all considered local
states.

For the {\it even states}, where ${\bf U_{\pm }^{{}}(-r)\!=U_{\pm }^{{}}(r)}$%
, the spectral system contains 7 independent variables. Those are components
of the impurity displacement, ${\bf U_{+}^{{}},}$ and radial projections of
the displacements of four chosen neighboring ions, $U_{s}.$ The elements of
the matrix in the corresponding spectral system can be written as integrals
of the Green's function elements, similar to Eq.~(33). Considering local
states near the gap's bottom, we again can retain in these integrals only
transverse terms and use the power expansions of non-singular factors of the
integrands. As it was shown above, 
\mbox{$\int^{}_{(\bf
k)}(ak)^2g^{-}_{\bot}\ll\int^{}_{(\bf k)}g^{-}_{\bot}$} \ due to the
dominant contribution of the small-$k$ region in these integrals. Therefore,
since the local states near the gap's bottom are, roughly,``made'' of long
wavelength polaritons, we can disregard the exponential factors in the
matrix elements of the spectral system [${\exp ({\pm i{\bf k\cdot n}}}%
)\approx 1$] and neglect in Eqs.~(21) differences between displacements of
identical ions within the elementary cell, ${\bf U_{-}^{{}}\approx
U_{-}^{{}}(n).}$ These approximations lead to the following spectral system
for even states: 
\begin{equation}
\left( 
\begin{array}{c}
{\bf U_{+}^{{}}} \\ 
{\bf U_{-}^{{}}}
\end{array}
\right) \approx \left( 
\begin{array}{cc}
{I^{+}} & {I^{\prime }} \\ 
{I^{\prime }} & {I^{-}}
\end{array}
\right) \left( 
\begin{array}{cc}
{\omega ^{2}\delta m\!-\!8\delta \beta /3} & \ {8\delta \beta /3} \\ 
{8\delta \beta /3} & {-8\delta \beta /3}
\end{array}
\right) \left( 
\begin{array}{c}
{\bf U_{+}} \\ 
{\bf U_{-}}
\end{array}
\right) .
\end{equation}
The corresponding spectral equation reads: 
\begin{equation}
\frac{8\delta \beta }{3}\left( I^{+}\!+\!I^{-}\!-\!2I^{\prime }\right)
\!-\!\omega ^{2}\delta mI^{+}\!+1=\frac{8\omega ^{2}\delta \beta \delta m}{3}%
\left( I^{+}I^{-}\!-\!I^{^{\prime }2}\right) .
\end{equation}
where all $I$-factors are straightforward to evaluate near the gap's bottom
[Appendix B]: 
\begin{equation}
I^{\pm }(\omega ^{2})\!=\frac{2}{3}\int_{({\bf k)}}^{{}}g_{\bot }^{\pm
}\approx \frac{2\left( ak_{0}\right) ^{2}}{3\pi \nu \sqrt{\omega
^{2}\!-\!\omega _{1}^{2}}}\frac{\mu \!-\!m_{\mp }}{m_{+}m_{-}},
\end{equation}
$$
I^{\prime }(\omega ^{2})\!=\frac{2}{3}\int_{({\bf k)}}^{{}}g_{\bot }^{\prime
}\approx \frac{2\left( ak_{0}\right) ^{2}}{3\pi \nu \sqrt{\omega
^{2}\!-\!\omega _{1}^{2}}}\frac{\mu }{m_{+}m_{-}}.\eqno(39') 
$$
The right-hand side of Eq.~(38) is proportional to a determinant of a
degenerate operator, the transverse propagator, and it must be equal to zero
identically. Substitution of Eqs.~(39,$39^{\prime }$) into Eq.~(38) shows
this explicitly and transforms the spectral equation of the even states into
the following one: 
\begin{equation}
\frac{2\left( ak_{0}\right) ^{2}}{3\pi \nu \mu \sqrt{\omega ^{2}\!-\!\omega
_{1}^{2}}}\left[ \frac{8\delta \beta }{3}-\omega _{0}^{2}\delta m\left( 
\frac{m_{-}}{M}\right) ^{2}\right] =1.
\end{equation}
One can check that here ions displacements satisfy the relationship, 
\mbox{$m^2_{-}{\bf
U_{+}}+m^2_{+}{\bf U_{-}}\!=\!0,$} corresponding to optical vibrations.

If we set $\delta\beta\!=\!0$ in Eq.~(40) it reproduces Eq.~(30) obtained
earlier for an isotope impurity. That equation has a solution only when $%
\delta m<0.$ In a general case even local states exist only if variations of
parameters satisfy the inequality: 
\begin{equation}
\delta \beta-\frac{3}{8}\omega^2_0 \delta m \left(\frac{m_{-}}{M}\right)^2>
0.
\end{equation}

Eqs.~(36,40) allow us to outline regions of the local states [Fig.3] on a
plane of impurity parameters, $\delta m,\delta\beta.$ \ Taking into account
obvious physical limitations, \mbox{$\delta m\ge-m_{+}$}, \ 
\mbox{$\delta
\beta\ge- \beta,$} one can see that odd states appear in the right upper
quadrant bounded by the lines $\delta m\!=\!-m_{+},\,\delta \beta\!=\!0.$
The region of even states is to the right of $\delta m\!=\!-m_{+}$ and above
the critical line, $\delta \beta\!= \!(3/8)\left(m_{-}/M\right)^2\omega^2_0
\delta m.$ All other areas of $\left(\delta m,\delta\beta\right)$-plane are
blocked for the polariton localization.

Since we deal with local states near the gap's bottom, our analysis is valid
in a vicinity of the critical line for even states, and close to $\delta m$%
-axis for odd states. When the impurity parameters move outside these
regions the frequencies of the corresponding local states move away from the
gap's bottom. The lines where the frequencies approach the top of the gap
establish the outer boundaries of the localization regions. Unlike the
situation near the gap's bottom, all terms of all spectral equations remain
finite when $\omega $ tends to $\omega _{2},$ what guarantees an existence
of limited localization regions. On the other hand, this makes it impossible
to do any rigorous analytical calculations. However, using the density of
states representation of the spectral equations, analogous to Eq.~(28), and
utilizing some trial functions to simulate densities of states in all bands,
one can obtain qualitatively reliable results. Our preliminary estimates
show that near the upper boundary of the gap local states are substantially
composed of transverse and longitudinal phonons. The balance between them
depends on widths of the corresponding bands and the polariton gap. The
contribution of the long wavelength transverse polaritons into these states
is proportional to $(v/c)^{3}$ and is negligible. Roughly, it is caused by
the fact that the density of states is inversely proportional to $v^{3}$ for
phonons and to $cv^{2}$ for long wavelength polaritons. A more detailed
analysis of states located far away from the gap's bottom will be presented
elsewhere.

\section{discussion and conclusions}

We have considered local polariton states in BCC ionic crystals. It was
assumed that the crystal anisotropy is weak and can be neglected in the long
wavelength limit. This approximation was proved to be self-consistent for
states located near the bottom of the polariton gap. We found two series of
local states, different in parity. The new states appear right at the bottom
of the polariton gap upon infinitesimally small variations of an impurity
parameters. This is in contrast with 3-D phonon systems where a lower
threshold for local states always exists.\cite{Lifshitz4,Maradudin,Figotin}
In Ref. \cite{Figotin} the general theorem regarding the presence of the
threshold for arising local states in bandgaps of periodic systems were
given. However, the proof assumed the finite values of density of states in
the entire band of pure system. We show that the singularity in the density
of states in the lower polariton band causes the absence of this threshold.
This singularity also provides that the states near the gap's bottom are
formed mostly by the long wavelength transverse polaritons. The local states
move toward the upper boundary of the gap upon increase of impurity
parameters $\delta m$ and/or $\delta \beta $.

We have outlined regions of new local polariton states on a plane of
impurity parameters. In a region where both types of states coexist, the odd
states precede the even ones. They appear first at the gap's bottom and
remain near it when the impurity parameters vary much longer than the even
states. Local states are a mixture of transverse phonons and photons.
Comparing amplitudes of the field, $E,$ and the crystal polarization, $P,$
one can make an estimate of the energy partitions of local polariton states.
Because the characteristic momentum, $k_{0},$ happened to be away from the
polariton resonance region, the ratio $E/P$ turns out to be of the order of $%
v_{\bot }^{\prime }/c.$ An account for the electronic polarization of ions
renormalizes $c$ and increases this ratio. More detailed analysis of these
aspects of the polariton local states will be done elsewhere. Condition (41)
can help in a search for compounds where local polariton states can be
observed.

The results of this paper were obtained within the harmonic approximation of
crystal dynamics. Phonon-phonon interaction caused by a lattice anharmonism
will affect this picture. Phonon-phonon scattering causes damping of all
elementary excitations and the local states as well. It leads to a
broadening of all spectral lines and it also washes out all sharp boundaries
in the initial excitation spectra. However, it seems physically evident that
the scale of this broadening is far below the typical width of phonon bands
and cannot change the topology of the initial band structure, or close the
polariton gap. If we dress up the elementary excitations and renormalize
their spectra, the maximum of the density of states in the lower polariton
band and the corresponding singularity will persist after the
renormalization since it has a topological origin. That, in turn, guarantees
the absence of a threshold for local states near the (renormalized) gap's
bottom at zero temperature. However, at non-zero temperature thermal
broadening of spectral lines will set a finite threshold for local states.

Another factor that leads to a threshold is crystal anisotropy. It provides
a difference between maxima of $\Omega _{-}^{2}({\bf k})$ in different
directions that causes the density of states in the lower polariton branch
to be finite everywhere. It should be emphasized, however, that properties
of the local polariton states remain quite different from the corresponding
properties of purely phonon local states. Particularly, one can see from our
results, that eigenfrequencies of local polaritons are much more sensible to
the crystal structure of the host materials.

\section*{Acknowledgments}

We wish to thank I.M. Vitebsky for useful discussion. This work was
supported by the NSF under grant No. DMR-9632789, by a CUNY collaborative
grant and PSC-CUNY grants.

\appendix

It is known that under any point transformation, ${\bf \hat{Q}}$, the
frequency of any normal mode remains invariant, $\omega^{2}({\bf \hat{Q}%
k)\!=\!\omega^{2}(k)}$, and the corresponding polarization vector transforms
as the follows: \mbox{${\bf e}(\bf\hat{Q}k)\!=\!{\bf\hat{Q}^{-1}e(k)}.$}
Because the Brillouin zone also maps exactly into itself, one can see that 
\begin{equation}
{\bf \hat{T}}\!=\! \int_{({\bf k)}}g[\omega^{2}({\bf k})]\,{\bf %
e(k)\!\otimes\! e(k)}\!=\! \int_{({\bf \hat{Q}k)}}g[\omega^{2}({\bf \hat{Q}k}%
)]\, {\bf e(\hat{Q}k)\!\otimes\!e(\hat{Q}k)}\!=\!{\bf \hat{Q}^{-1}\hat{T}%
\hat{Q}}.
\end{equation}
Therefore, any tensor of this type is an invariant of the point group of a
crystal. In the cubic system any group-invariant operator must be trivial, $%
{\bf \hat{T}}\!=\!t\,{\bf \hat{I}},$ since the group contains non-collinear
axis of different order. Calculating the trace of the operator, one can
find: 
\begin{equation}
{\bf \hat{T}}\!=\!1/3\int_{({\bf k)}}g[\omega^{2}({\bf k)] \left[e(k)\cdot
e(k)\right]\:\,\hat{I},}
\end{equation}
where ${\bf \hat{I}}$ is the identity operator.

The density of states in the lower polariton band, $\rho_{-}(\varepsilon),$ 
is defined at the complex $\varepsilon$-plane, cut from $0$ to $\omega_1^{2}.
$ The contour of integration, $C,$ in Eq.~(26) runs along the upper side of
the cut (it corresponds to integration over $k<k_0$), turns around its right
edge and then, for the integration over $k>k_0,$ returns along the lower
side of the cut to some point, $\omega_1^{^{\prime}2},$ fixed by the bottom
of the optical photon band.

As it was discussed in the text, the leading part of $I(\omega^2)$ at $%
\Delta(\omega)\!=\!\omega^2\!-\!\omega_1^2\ll\omega_1^2,$ comes from the
region close to the right edge of the cut, $\varepsilon\!=\!\omega_1^{2}.$
Therefore, evaluating Eq.~(26), we can use there the asymptote of the
density of states given by Eq.~(27). In addition, as it follows from
Eq.~(25), $k\!-\!k_0\!=\mp z^{1/2}/a\nu$ near the right edge of the cut at
its upper and lower sides, respectively. Taking this into account when
expanding $F(k)$ around $k_0$ in Eq.~(26), we obtain: 
\begin{equation}
I^{+}(\omega^2)\!=\!\int_{0}^{\omega^2_1}\frac{2\left[F_0\!-\!F^{\prime}_0 
\sqrt{z}/{a\nu}\!+\cdots\right]dz} {3\sqrt{z}\left(\Delta\!+\!z\right)}
\!+\! \int_{0}^{\omega^2_1\!-\!\omega^{^{\prime}2}_1}\frac{%
2\left[F_0\!+\!F^{\prime}_0 \sqrt{z}/{a\nu}\!+\cdots\right]dz} {3\sqrt{z}%
\left(\Delta\!+\!z\right)}.
\end{equation}
A rescaling $z\!\rightarrow\!z\Delta$ shows that contributions in $%
I^{+}(\omega^2)$ from all terms of expansion of $F(k)$ tend to zero in the
limit $\Delta\!\rightarrow\! +0,$ except for the first two terms. Further
elementary integration gives the singular part of the considered integral: 
\begin{equation}
I^{+}(\omega^2)=\frac{2(ak_0)^2 F(k_0)}{3\pi\nu\Delta^{1/2}}=\frac{2(ak_0)^2 
}{3\pi\nu\Delta^{1/2}} \frac{\left(\omega^2\!-\!c^2k^2_0\right)
\left[\gamma_{\bot}(k_0)\!+\!f_{\bot}(k_0)\!-\!m_{-}\omega^2\right]} {
m_{+}m_{-}\left[\omega^2\!-\!\Omega_{+}^{2}(k_0)\right]
\left[\omega^2\!-\!\omega_{\bot}^{2}(k_0)\right]}.
\end{equation}
This result is asymptotically exact since all omitted terms are regular at $%
\Delta\!\rightarrow\!+0.$ Using estimations made in Section 2. and taking
into account conditions, $ak_0\ll 1,$ \ $v^{^{\prime}2}_{\bot}k^2_0\ll%
\omega_0^2,$ \ $c^2k^2_0\gg\omega_0^2$ \ and \ $\omega\approx\omega_1\approx%
\omega_0,$ \ it is straightforward to obtain: 
\begin{equation}
I^{+}(\omega^2)\approx- \frac{2\left(ak_0\right)^2}{3\pi\nu\sqrt{%
\omega^2\!-\!\omega_1^2}} \frac{m_{-}}{Mm_{+}}
\end{equation}
which immediately leads to Eq.~(28).

In a similar way one can obtain Eqs.~(37), as well as, calculate the
integral in Eqs.~(33): 
\begin{equation}
\int^{}_{({\bf k)}}(ak)^2g^{-}_{\bot}=\frac{2(ak_0)^4 }{3\pi\nu\Delta^{1/2}} 
\frac{\left(\omega^2\!-\!c^2k^2_0\right)
\left[\gamma_{\bot}(k_0)\!+\!f_{\bot}(k_0)\!-\!m_{+}\omega^2\right]} {%
m_{+}m_{-}\left[\omega^2\!-\!\Omega_{+}^{2}(k_0)\right]
\left[\omega^2\!-\!\omega_{\bot}^{2}(k_0)\right]}\approx -\frac{%
2\left(ak_0\right)^4\omega_0^2} {3\pi\nu\sqrt{\omega^2\!-\!\omega_1^2}}\,%
\frac{m_{+}}{Mm_{-}}
\end{equation}

Eq.~(30) with the matrix $M_{ss^{\prime}}$ defined by Eq.~(31) can be
rewritten as follows: 
\begin{equation}
\left( 
\begin{array}{cccc}
{M_{11}\!-\!M_{13}\!-\!\mu} & {0} & {0} & {0} \\ 
{0} & {M_{11}\!-\!M_{13}\!-\!\mu} & {0} & {0} \\ 
{M_{13}} & {M_{12}} & {M_{11}\!+\!M_{13}\!-\!\mu} & {2M_{12}} \\ 
{M_{12}} & {M_{13}} & {2M_{12}} & {M_{11}\!+\!M_{13}\!-\!\mu}
\end{array}
\right) \left( 
\begin{array}{c}
{U_1} \\ 
{U_2} \\ 
{U_3\!-\!U_1} \\ 
{U_4\!-\!U_2}
\end{array}
\right)=0.
\end{equation}
which has three eigenvalues given by Eqs.~(32).

For $\mu=\mu_{\pm}$ this equation gives $U_1=U_2=0$ and $U_3=\pm U_4.$
Recalling that $U_s={\bf n}\cdot{\bf U}({\bf n}_s)$ and the states we
consider here are antisymmetric, one can see that in $\mu_{+}$-mode the
elementary cell inclosing a defect experiences ``rhombic'' deformations. In $%
\mu_{-}$-mode, since the ion motion is anti-phased, it produces
``tetragonal'' deformations. For $\mu=\mu_0$ Eq.~(C1) leaves $U_1$ and $U_2$
independent, whereas, $U_3=U_1/2$ and $U_4=U_2/2.$ It results in combined,
``tetra-rhombic''deformations of the elementary cell in these modes.

To obtain the explicit expressions of the eigenvalues, $\mu ,$ we have to
calculate the matrix elements $M_{ss^{\prime }}.$ When the frequency of the
considered states is close to $\omega _{1}$ we can retain only first terms
in expansions of $\sin {}$-factors in Eq.~(29). This approximation is
self-consistent, because the region of small wave vectors is the major
contribution in Eq.~(29). Carrying out the calculations, one needs to know
tensors of two types: \mbox{$\int_{\bf k}f(k)k^{\alpha}k^{\beta}$} and 
\mbox{$\int_{\bf
k}f(k)k^{\alpha}k^{\beta}k^{\sigma}k^{\beta},$} where $f(k)$ is an arbitrary
invariant function. From Appendix A it immediately follows that, $\int_{{\bf %
k}}f(k)k^{\alpha }k^{\beta }=(1/3)\delta ^{\alpha \beta }\int_{{\bf k}%
}f(k)k^{2}.$ The second tensor is obviously totally symmetric, therefore: 
\begin{equation}
\int_{{\bf k}}f(k)k^{\alpha }k^{\beta }k^{\sigma }k^{\beta }=I\left( \delta
^{\alpha \beta }\delta ^{\sigma \beta }+\delta ^{\alpha \sigma }\delta
^{\beta \beta }+\delta ^{\alpha \beta }\delta ^{\beta \sigma }\right) \int_{%
{\bf k}}f(k)k^{4},
\end{equation}
where summation over $\alpha =\beta $ \ and \ $\sigma =\beta $ \ defines the
numerical factor, $I=1/15.$ Using these results in $M_{ss^{\prime }}$ \ and
retaining there the transverse terms only, we obtain: 
\begin{equation}
M_{ss^{\prime }}\approx \frac{1}{3}\left\{ ({\bf n}_{s}{\bf n}_{s^{\prime
}})^{2}-\frac{n^{4}+2({\bf n}_{s}{\bf n}_{s^{\prime }})^{2}}{5}\right\}
\int_{({\bf k)}}^{{}}k^{2}g_{\bot }^{-},
\end{equation}
that, in turn, leads to Eqs.~(33).

\pagebreak

\subsection*{Figure Captions}

Fig. 1. Elementary cell of a body centered cubic lattice. Here $a$ is a
lattice parameter, indexes $s,s^{\prime }$ numerate ions within the
near-neighbor shell, $\pm {\bf n}_{s}$ are radius-vectors of the ions.

Fig. 2. Phonon-polariton dispersion curves in the isotropic approximation.
Here lines $\omega =\Omega _{\pm }(k)$ present the lower and upper polariton
branches, $\Omega _{\bot }(k)$ and $\omega _{+\Vert }(k)$ are transverse and
longitudinal optical phonon branches, $\omega _{\bot }(k)$ and $\omega
_{-\Vert }(k)$ are the transverse and longitudinal acoustic phonon branches,
respectively. The horisontal lines $\omega _{0}$ and $\omega _{3}$ are the
upper and lower boundaries of the transverse optical phonon band, $\sqrt{%
\omega _{0}^{2}+d^{2}}$ and $\omega _{2}$ are the upper and lower boundaries
of the longitudinal optical phonon band, respectively. The polariton gap is
bounded by lines $\omega _{1}$ and $\omega _{2}.$ The wave number $k_{0}$
corresponds to the maximum of the lower polariton branch.

The figure presents qualitative features of phonon-polariton spectra in the
long wavelength region and does not reflect an actual scale of wave
velocities.

Fig. 3. Regions of existence of local polariton states. $\delta m$ is the
difference between masses of the impurity and a host atom, $\delta \beta $
is the impurity induced deviation of the elastic constant within the
near-neighbor shell, $m_{+}$ is the mass of a host ion in the ``positive''
sublattice, $\beta $ is the elastic constant of a near-neighbor interaction
in a perfect crystal. Odd local polariton states exist in the region bound
by the solid lines $\delta m\!=\!-m_{+},\,\delta \beta\!=\!0.$ The region of
even local states is to the right of the solid line $\delta m\!=\!-m_{+}$
and above the dashed line, $\delta \beta\!=
\!(3/8)\left(m_{-}/M\right)^2\omega^2_0 \delta m .$

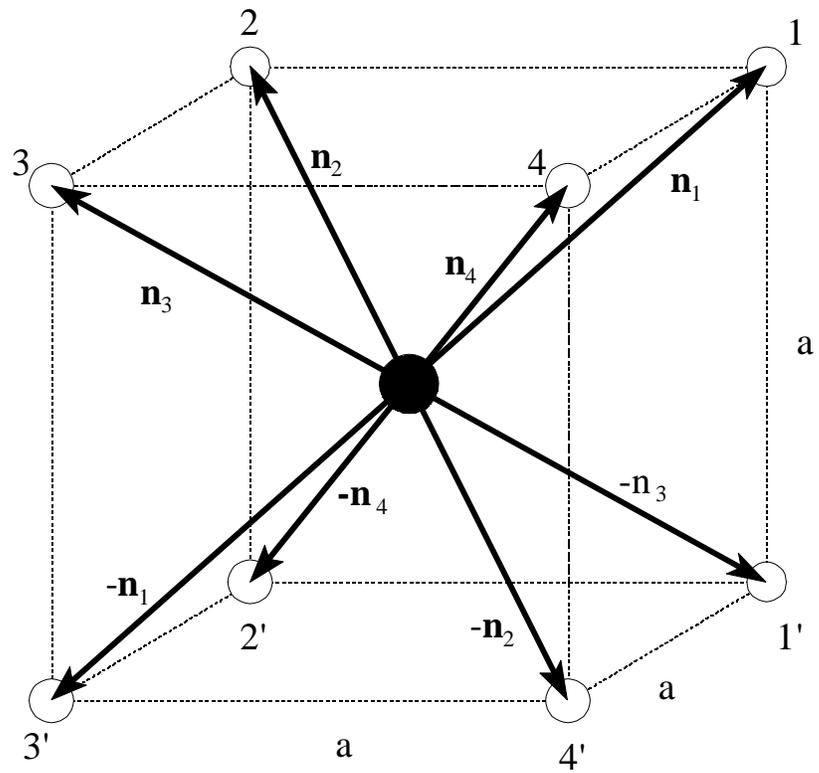
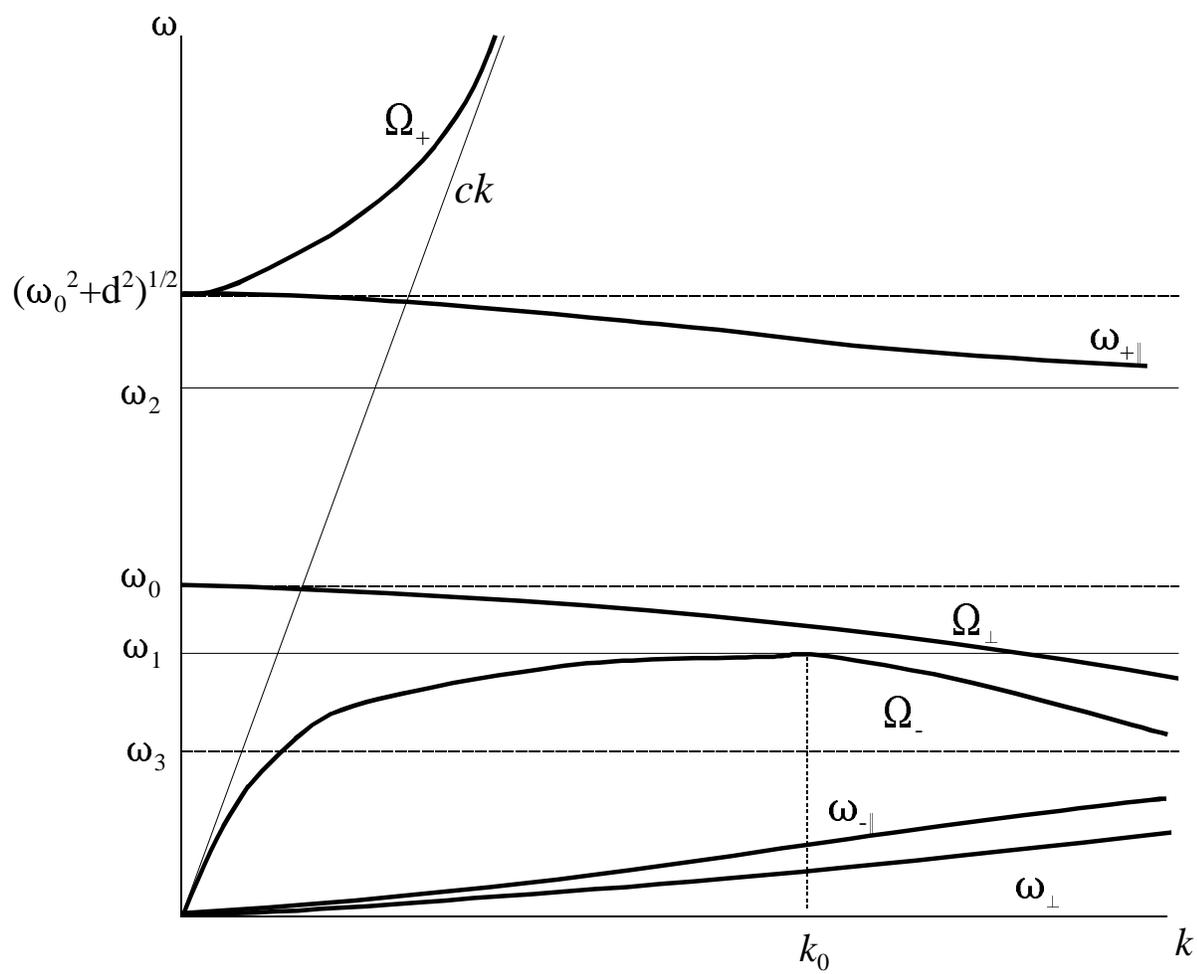
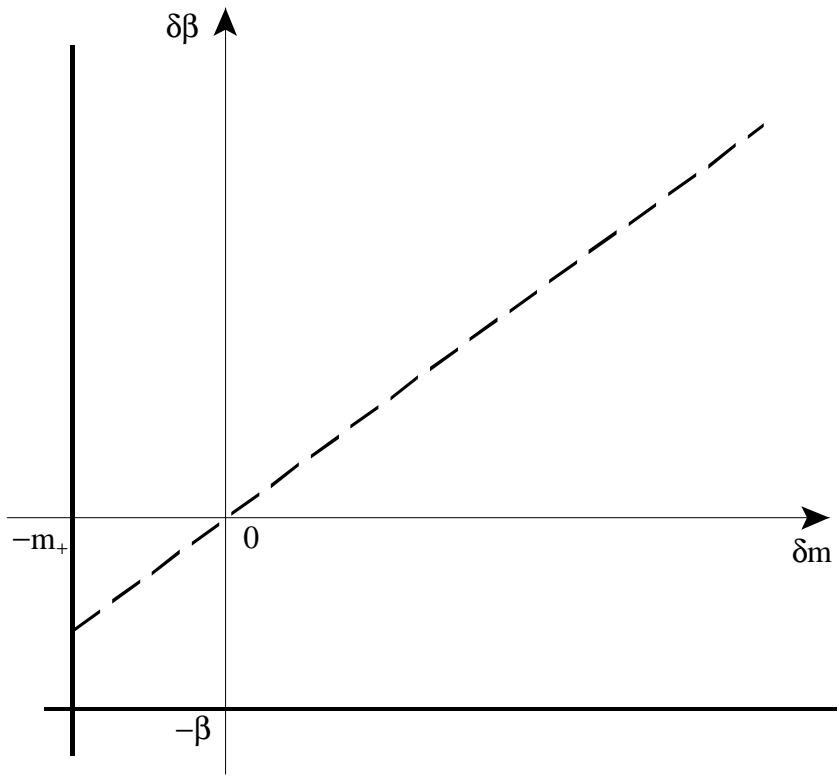
\end{document}